\documentclass[letter,10pt,conference]{IEEEtran}
\usepackage{pifont}
\usepackage{amsfonts}
\usepackage[T1]{fontenc}
 \usepackage[bookmarks=false]{hyperref}
\ifCLASSINFOpdf
\else
\fi
\usepackage{relsize}

\usepackage{centernot}

\usepackage{cite}
\usepackage{amsfonts}
\usepackage{mathrsfs}
\usepackage{bbm}
\usepackage{pifont}
\usepackage{amsfonts}
\usepackage{amsmath, amsthm}
\usepackage{tabularx}
\usepackage{listings}
\usepackage{graphicx}
\usepackage{float}
\usepackage{amssymb}
\usepackage{array}
\usepackage{fancyhdr}
\usepackage{cases}
 \usepackage{supertabular}

\usepackage{url} 
\usepackage{fancyhdr}

\usepackage{psfrag}
\usepackage{color}
\usepackage{bbding}
\newcommand{\bcap} {\hspace{2pt} \mathlarger{\cap}
\hspace{2pt}}

\newcommand {\C} {{\rm I\kern-5.5pt C}}

\newcommand{\bP}[1]{{\mathbb{P}}\left[{#1}\right]}

\def\centerhack#1{\hbox to 0pt{\hss\footnotesize #1\hss}}
\def\centerhackn#1{\hbox to 0pt{\hss #1\hss}}
\def\dchack#1{\vbox to 0pt{\vss{\hbox to 0pt{\hss#1\hss}}\vss}}

\newtheorem{lem}{Lemma}
\newtheorem{thm}{Theorem}

\newtheorem{rem}{Remark}
\newtheorem{cor}{Corollary}

\newtheorem*{proposition1.1}{Proposition 1.1}
\newtheorem*{proposition1.2}{Proposition 1.2}
\newtheorem*{proposition1.3}{Proposition 1.3}
\newtheorem*{proposition2.1}{Proposition 2.1}
\newtheorem*{proposition2.2}{Proposition 2.2}
\begin{document}

\title{\LARGE On Topological Properties of Wireless Sensor Networks under
the $q$-Composite Key Predistribution Scheme with On/Off Channels}

\author{ \IEEEauthorblockN{Jun Zhao}
\IEEEauthorblockA{CyLab and Dept.
of ECE \\
Carnegie Mellon University \\
Pittsburgh, PA 15213\\
Email: junzhao@junzhao.info} \and \IEEEauthorblockN{Osman Ya\u{g}an}
\IEEEauthorblockA{CyLab and Dept.
of ECE\\
Carnegie Mellon University \\
Moffett Field, CA 94035\\
Email: oyagan@ece.cmu.edu} \and \IEEEauthorblockN{Virgil Gligor}
\IEEEauthorblockA{CyLab and Dept.
of ECE \\
Carnegie Mellon University \\
Pittsburgh, PA 15213\\
Email: gligor@cmu.edu}}

%
%\author{\IEEEauthorblockN{Jun Zhao, Osman Ya\u{g}an and Virgil Gligor}
%\IEEEauthorblockA{\\CyLab and Dept.
%of ECE \\
%Carnegie Mellon University \\ \{junzhao, oyagan,
%virgil\}@andrew.cmu.edu}}

%\author{\IEEEauthorblockN{Jun Zhao, Osman Ya\u{g}an and Virgil Gligor}
%\IEEEauthorblockA{{\tiny~\vspace{-7pt}}\\CyLab and Dept.
%of ECE \\
%Carnegie Mellon University \\ \{junzhao, oyagan,
%virgil\}@andrew.cmu.edu\vspace{-7pt}}}

%\author{\IEEEauthorblockN{~}
%\IEEEauthorblockA{~\\
%~ \\ ~}}

\maketitle

\makeatletter
\let\old@ps@headings\ps@headings
\let\old@ps@IEEEtitlepagestyle\ps@IEEEtitlepagestyle
\def\confheader#1{%
  % for all pages except the first
  \def\ps@headings{%
    \old@ps@headings%
    \def\@oddhead{\strut\hfill#1\hfill\strut}%
    \def\@evenhead{\strut\hfill#1\hfill\strut}%
  }%
  % for the first page
  \def\ps@IEEEtitlepagestyle{%
    \old@ps@headings%
    \def\@oddhead{\strut\hfill#1\hfill\strut}%
    \def\@evenhead{\strut\hfill#1\hfill\strut}%
  }%
  \ps@headings%
} \makeatother

\confheader{\large{\color{red}Best Student Paper Finalist}~in 2014 IEEE International Symposium on Information Theory (ISIT)\\The full }

\lhead{}

\rhead{}

\thispagestyle{fancy}
\pagestyle{fancy}

\markboth{a}{b}

\fancyhead[C]{\large{\color{red}Best Student Paper Finalist}~in 2014 IEEE International Symposium on Information Theory (ISIT)\\The detailed proofs are presented in the full version
\cite{QcompTech14} (CMU Technical Report
  CMU-CyLab-14-002).}

%\markboth{} {\large This paper has been accepted at IEEE ISIT 2014
%to be held in June--July 2014.}

\begin{abstract}
\boldmath The $q$-composite key predistribution scheme \cite{adrian}
is used prevalently for secure communications in large-scale
wireless sensor networks (WSNs). Prior work \cite{yagan_onoff, ISIT,
ZhaoYaganGligor} explores topological properties of WSNs employing
the $q$-composite scheme for $q=1$ with unreliable communication
links modeled as independent on/off channels. In this paper, we
investigate topological properties related to the node degree in
WSNs operating under the $q$-composite scheme and the on/off channel
model. Our results apply to general $q$ and are stronger than those
reported for the node degree in prior work even for the case of $q$
being $1$. Specifically, we show that
 the number of nodes with certain degree is asymptotically equivalent in distribution to a Poisson random variable, present the asymptotic probability
distribution for the minimum degree of the network, and establish
the asymptotically exact probability for the property that the
minimum degree is at least an arbitrary value. Numerical experiments
confirm the validity of our analytical findings.

\end{abstract}

%\vspace{10pt}
%
%\begin{IEEEkeywords}
%key predistribution, node degree, random graph, random intersection
%graph, random key graph, security, topological properties, wireless
%sensor networks.
% \end{IEEEkeywords}


\vspace{-10pt}

%$p_{s,q}$ is given in Proposition 2 of \cite{Rybarczyk}? or
%somewhere else?

\section{Introduction}

The basic key predistribution scheme of Eschenauer and Gligor
\cite{virgil} has been recognized as a typical solution to secure
communication in wireless sensor networks (WSNs) and studied
extensively in the literature over the last decade [1]--[11]. %[1]--[3], [6],
%[9]--[14] \cite{Zhu:2006:LES:1218556.1218559}.
 The idea is that cryptographic keys are assigned before deployment to ensure secure
sensor-to-sensor communications.

The $q$-composite key predistribution scheme proposed by Chan
\emph{et al.} \cite{adrian} as an extension of the basic
Eschenauer--Gligor scheme \cite{virgil} (the $q$-composite scheme in
the case of $q=1$) has received much interest \cite{Rybarczyk,bloznelis2013,Assortativity,DBLP:conf/waw/BradonjicHHP10,Bloznelis201494,ANALCO,Perfectmatchings}
 since its introduction.

 The $q$-composite scheme works as follows. For a WSN with $n$
sensors, prior to deployment, each sensor is independently assigned
$K_n$ different keys which are selected \emph{uniformly at random}
from a pool of $P_n$ keys.  Then two sensors establish a
\emph{secure} link in between after deployment \emph{if and only if}
they share at least $q$ key(s) \emph{and} the physical link
constraint between them is satisfied. $P_n$ and $K_n$ are both
functions of $n$ for generality, with the natural condition $1 \leq
K_n \leq P_n$. Examples of physical link constraints include the
reliability of the transmission channel and the distance between two
sensors close enough for communication. The $q$-composite scheme
with $q\geq 2$ outperforms the basic Eschenauer--Gligor scheme with
$q=1$ in terms of the strength against small-scale network capture
attacks while trading off increased vulnerability in the face of
large-scale attacks \cite{adrian}.

 In this
paper, we investigate topological properties related to
node\footnote{A sensor is also referred to as a node.} degree in
WSNs employing the $q$-composite key predistribution scheme with
general $q$ under the \emph{on}/\emph{off} channel model as the
physical link constraint comprising independent channels which are
either \emph{on} or \emph{off}. The degree of a node $v$ is the
number of nodes having secure links with $v$; and the minimum (node)
degree of a network is the least among the degrees of all nodes.
Specifically, we demonstrate that the number of nodes with certain degree is asymptotically equivalent in distribution to a Poisson random variable,
establish the asymptotic probability distribution for the minimum
degree of the network, and derive the asymptotically exact
probability for the property that the minimum degree is no less than
an arbitrary value. Ya\u{g}an \cite{yagan_onoff} and we
\cite{ISIT,ZhaoYaganGligor} consider the WSNs with $q=1$ and show
results for several topological properties, yet results about node
degree in these prior work are weaker than our analytical findings
even when the general $q$ is set as $1$.

Our approach to the analysis is to explore the induced random graph
models of the WSNs. As will be clear in Section
\ref{sec:SystemModel}, the graph modeling a WSN under the
$q$-composite scheme and the on/off channel model is an intersection
of two distinct types of random graphs. It is the intertwining
\cite{yagan_onoff,DBLP:journals/jgt/BollobasS11,JZIS14,Penrose2013}
of these two graphs that makes our analysis challenging.

We organize the rest of the paper as follows. Section
\ref{sec:SystemModel} describes the system model in detail.
Afterwards, we present and discuss the results in Section
\ref{sec:res}. Subsequently, we provide numerical experiments in
Section \ref{sec:expe} to confirm our analytical results, whereas
Section \ref{related} is devoted to relevant results in the
literature. Next, we conclude the paper and identify future research
directions in Section \ref{sec:Conclusion}.

\section{System Model}
\label{sec:SystemModel}

We elaborate the graph modeling of a WSN with $n$ sensors, which
employs the $q$-composite key predistribution scheme and works under
the {on/off} channel model. We use a node set $\mathcal {V} = \{v_1,
v_2, \ldots, v_n \}$ to represent the $n$ sensors. For each node
$v_i \in \mathcal {V}$, the set of its $K_n$ different keys is
denoted by $S_i$, which is uniformly distributed among all
$K_n$-size subsets of a key pool of $P_n$ keys.

The $q$-composite key predistribution scheme is
 modeled by a uniform $q$-intersection graph denoted by\footnote{Many papers
\cite{Rybarczyk,bloznelis2013,Assortativity,DBLP:conf/waw/BradonjicHHP10,Bloznelis201494,ANALCO,Perfectmatchings}
in the literature use $s$ instead of $q$ so we have uniform
$s$-intersection graph $G_s(n, K_n, P_n)$. This work uses $q$
following the $q$-composite key predistribution scheme
\cite{adrian}.} $G_q(n,K_n,P_n)$, which is defined on the node set
$\mathcal{V}$ such that any two distinct nodes $v_i$ and $v_j$
sharing at least $q$ key(s) (an event denoted by $\Gamma_{ij}$) have
an edge in between. Clearly, $\Gamma_{ij}$ equals event $\big[ |S_i
\bcap S_j | \geq q \big]$, where $|A|$ with $A$ as a set means the
cardinality of $A$.

Under the {on/off} channel model, each node-to-node channel is
independently {\em on} with probability $p_n $ and {\em off} with
probability $(1-p_n)$, where $p_n$ is a function of $n$ with
$0<p_n\leq 1$. Denoting by ${C}_{i j}$ the event that the channel
between distinct nodes $v_i$ and $v_j$ is {\em on}, we have
$\bP{C_{ij}} = p_n$, where $\mathbb{P}[\mathcal {E}]$ denotes the
probability that event $\mathcal {E}$ happens, throughout the paper.
The {on/off} channel model is represented by an Erd\H{o}s-R\'enyi
graph $G(n, p_n)$ \cite{citeulike:4012374} defined on the node set
$\mathcal{V}$ such that $v_i$ and $v_j$ have an edge in between if
event $C_{ij}$ occurs.

Finally, we denote by $\mathbb{G}_q (n, K_n, P_n,
p_n)$ the underlying graph of the $n$-node WSN operating under the
$q$-composite key predistribution scheme and the on/off channel
model. We often write $\mathbb{G}_q$ rather than $\mathbb{G}_q(n,
K_n, P_n, p_n)$ for notation brevity. Graph
$\mathbb{G}_q$ is defined on the node set
$\mathcal{V}$ such that there exists an edge between nodes $v_i$ and
$v_j$ if and only if events $\Gamma_{ij}$ and $C_{ij}$ happen at the
same time. We set event $E_{ij} : = \Gamma_{ij} \cap C_{ij}$.
%\begin{equation}
%E_{ij} = \Gamma_{ij} \cap C_{ij},\textrm{ for }1\leq i < j \leq n,
%\label{eq:E_is_K_cap_C_oy}
%\end{equation}
Then $\mathbb{G}_q$ can be seen as the intersection
of $G_q(n, K_n, P_n)$ and $G(n, p_n)$, i.e.,
\begin{equation}
\mathbb{G}_q  = G_q(n, K_n, P_n) \cap G(n, p_n).
\nonumber
%\label{eq:G_on_is_RKG_cap_ER_oy}
\end{equation}
%Note that the parameters of $\mathbb{G}_q $ include
%$n, K_n, P_n, p_n$ and $q$.

We define $p_{s,q} $ as the probability that two different nodes
share at least $q$ key(s) and $p_{e,q} $ as the probability that two
distinct nodes have a secure link in $\mathbb{G}_q$. Clearly,
$p_{s,q} $ and $p_{e,q}$ are the edge probabilities in graphs
$G_q(n, K_n, P_n)$ and $\mathbb{G}_q$, respectively. $p_{s,q} $ and
$p_{e,q}$ both depend on $K_n, P_n$ and $q$, while $p_{e,q}$ also
depends on $p_n$. By definition, $p_{s,q}$ is determined through
\begin{align}
p_{s,q} & =  \mathbb{P} [\Gamma_{i j} ] = \sum_{u=q}^{K_n}
  \mathbb{P}[|S_{i} \cap S_{j}| = u] , \label{psq1}
\end{align}
where it is shown \cite{QcompTech14,Rybarczyk} that
\begin{align}
& \mathbb{P}[|S_{i} \cap S_{j}| = u] \nonumber \\ \quad & =
 \begin{cases} \frac{\binom{K_n}{u}\binom{P_n-K_n}{K_n-u}}{\binom{P_n}{K_n}} ,
&\textrm{for } \max\{0,2K_n-P_n\} \leq u \leq K_n , \\ 0,
&\textrm{otherwise}.
\end{cases} \label{psq2}
\end{align}
From $E_{ij} = \Gamma_{ij} \cap C_{ij}$ and the independence of
${C}_{i j} $ and $ \Gamma_{i j} $, we obtain
\begin{align}
{p_{e,q}}  & =  \mathbb{P} [E_{i j} ]  =  \mathbb{P} [{C}_{i j} ]
\cdot \mathbb{P} [\Gamma_{i j} ] =  p_n \cdot
p_{s,q}. \label{eq_pre}
\end{align}

\section{The Results and Discussion} \label{sec:res}

We present and discuss our results in this section. Throughout the
paper, $q$ is an arbitrary positive integer and does not scale with
$n$;
%$\mathbb{N}_0 $ stands for the set of all positive integers;
%$\mathbb{R}$ is the set of all real numbers;
 $e$ is the base of the
natural logarithm function, $\ln$. All limits are understood with $n
\to
  \infty$.  %The term ``for all $n$
%sufficiently large'' means ``for any $n \geq N$, where $N \in
%\mathbb{N}_0$ is selected appropriately''.
 We use the standard
asymptotic notation $o(\cdot), \omega(\cdot), O(\cdot),
\Theta(\cdot), \sim$. In particular, for two sequences $f_n$ and
$g_n$, $f_n \sim g_n$ signifies $\lim_{n \to
  \infty}({f_n}/{g_n})=1$; namely, $f_n$
  and $g_n$ are asymptotically equivalent.

\subsection{Results of Graph
$\mathbb{G}_q$}

We detail the results of graph $\mathbb{G}_q$ in Theorem \ref{thm:exact_qcomposite2} and
Corollary \ref{cor:exact_qcomposite} below. The detailed proofs of
Theorem \ref{thm:exact_qcomposite2} and Corollary
\ref{cor:exact_qcomposite} can be found in the full version
\cite{QcompTech14} and are omitted here owing to the space
limitation. The basic idea is to use the method of moments
\cite{JansonLuczakRucinski7}.
%
%Denoting by $\delta$ the minimum node degree of graph
%$\mathbb{G}_q$, we detail the results of
%$\mathbb{G}_q$ below.
%

%
%
%The proof of Theorem \ref{thm:exact_qcomposite} detailed in Appendix
%similar to that of Theorem \ref{thm:exact}, with necessary
%modifications.
%
%See our technical report.
%
%\subsubsection{The Distributions of the
%Minimum Node Degree and the Connectivity in Graph
%$\mathbb{G}_q$}~

\begin{thm} \label{thm:exact_qcomposite2}
For graph $\mathbb{G}_q$ under $ K_n =
\omega(1)$  and $\frac{{K_n}^2}{P_n} = o(1)$, the following
properties (a) and (b) hold.

\textbf{(a)} If
\begin{align}
p_{e, q} & = \Theta\bigg( \frac{\ln  n}{n} \bigg),  \label{peq1}
\end{align}
then for $h = 0,1,2,\ldots$, it follows that $\phi_h$ denoting the number of nodes with degree $h$, 
is asymptotically equivalent in distribution to a Poisson random variable
with mean $\lambda_{n,h} : =  n (h!)^{-1}(n p_{e,q})^h e^{-n p_{e,q}}$. In other words, with $\textrm{Po}(\lambda_{n,h} )$ denoting a Poisson random variable
with mean $\lambda_{n,h} $, it
holds that
\begin{align}
\mathbb{P}[\phi_h = i]  & \sim \mathbb{P}[ \textrm{Po}( \lambda_{n,h} ) = i], \quad\textrm{for } i=0,1,2,\ldots. \nonumber
\end{align}
 
\textbf{(b)} If for some integer $\ell$ and some sequence $\alpha_n$
satisfying
\begin{align}
-1  &  < \liminf_{n \to \infty} \frac{\alpha_n}{\ln \ln n} \leq
\limsup_{n \to \infty} \frac{\alpha_n}{\ln \ln n}< 1,  \nonumber
\end{align}
it holds that
\begin{align}
p_{e, q} & = \frac{\ln  n + {(\ell-1)} \ln \ln n + {\alpha_n}}{n},
\label{peq1sbsc}%\label{newpeq}
\end{align}
%Equations (\ref{e1}) and (\ref{e2}) are determined by finding $\ell$
%and $\alpha_n$ with $-\frac{1}{2} \ln \ln n \leq \alpha_n <
%\frac{1}{2} \ln \ln n$ such that $ $.
%
% Defining $\ell$ and $\alpha_n$ by
then defining $\delta$ as the minimum degree of
$\mathbb{G}_q$, we obtain:\\ for $\ell \leq 0$, it
follows that as $n \to \infty$,
\begin{align}
\hspace{-3pt}\begin{cases} \delta =  0\textrm{ with a probability approaching to }1, \\
\delta > 0\textrm{ with a probability going to }0; \nonumber
\end{cases}
\end{align}
and for $\ell > 0$, properties (b1)--(b4) below hold.
\begin{itemize}
\item [\emph{(b1)}] $(\delta \neq  \ell)\cap (\delta \neq  \ell-1)$ \textrm{ with
a probability going to }0 as $n \to \infty$;
\item [\emph{(b2)}] if $\lim\limits_{n \to \infty} \alpha_n = \alpha ^* \in (-\infty, \infty)$, then as $n \to
\infty$,
\begin{align}
\begin{cases} \delta =  \ell \textrm{ with a probability converging to }e^{- \frac{e^{-\alpha ^*}}{(k-1)!}}, \\
\delta =   \ell - 1\textrm{ with a probability tending to }\Big(
\hspace{-1pt} 1 \hspace{-2pt} - \hspace{-2pt} e^{- \frac{e^{-\alpha
^*}}{(k-1)!}} \hspace{-1pt} \Big);\nonumber
\end{cases}
\end{align}
  \item [\emph{(b3)}] if $ \lim\limits_{n \to \infty} \alpha_n = \infty$, then as $n \to
\infty$,
  \begin{align}
\hspace{-3pt}\begin{cases} \delta =  \ell\textrm{ with a probability approaching to }1, \\
\delta \neq  \ell\textrm{ with a probability going to }0;
\end{cases}\nonumber \hspace{20pt}\textrm{and}
\end{align}
  \item [\emph{(b4)}] if $ \lim\limits_{n \to \infty} \alpha_n = - \infty$, then as $n \to
\infty$,
  \begin{align}
\hspace{-27pt}\begin{cases} \delta =  \ell - 1\textrm{ with a probability tending to }1, \\
\delta \neq   \ell - 1\textrm{ with a probability converging to }0.
\end{cases}\nonumber
\end{align}
\end{itemize}

\end{thm}

%\begin{align}
%\ell : = \bigg\lfloor \frac{np_{e, q} - \ln n + (\ln \ln n) / 2}{\ln
%\ln n} \bigg\rfloor + 1, \label{e1}
%\end{align}
%and
% \begin{align}
%\alpha_n : = np_{e, q} - \ln n - (\ell-1)\ln\ln n, \label{e2}
%\end{align}

\begin{rem}
Property (b) of Theorem \ref{thm:exact_qcomposite2}
presents the asymptotic probability distribution for the minimum
degree of the network.

%
%, where an asymptotic Poisson distribution of a variable $\nu_n$
%means that there exists another variable $\mu_n $ following a
%Poisson distribution such that $\mathbb{P}[\nu_n = i] \sim
%\mathbb{P}[\mu_n = i]$ (i.e., $\lim\limits_{n \to
%  \infty} \frac{\mathbb{P}[\nu_n = i]}{\mathbb{P}[\mu_n = i]} =1$) for any non-negative integer $i$.
%

\end{rem}

%\begin{rem}
%
%Equations (\ref{e1}) and (\ref{e2}) are determined by finding $\ell$
%and $\alpha_n$ with $-\frac{1}{2} \ln \ln n \leq \alpha_n <
%\frac{1}{2} \ln \ln n$ such that $ p_{e, q} = \frac{\ln  n +
%{(\ell-1)} \ln \ln n + {\alpha_n}}{n}$.
%
%
%\end{rem}

We present a corollary of Theorem
\ref{thm:exact_qcomposite2} below. The corollary is established with the help of
a graph coupling argument \cite{zz} (see the full version
\cite{QcompTech14} for the detailed proof).

\begin{cor}\label{cor:exact_qcomposite}
For graph $\mathbb{G}_q$ under $ K_n =
\omega(1)$ and $\frac{{K_n}^2}{P_n} = o(1)$, with some sequence $\beta_n$ defined by
\begin{align}
p_{e, q} & =  \frac{\ln  n + {(k-1)} \ln \ln n + {\beta_n}}{n} 
 \label{peq2}
\end{align}
for some positive integer $k$, and with $\delta$ denoting the minimum
degree of $\mathbb{G}_q$, it holds that
\begin{align}
 \lim_{n \to \infty} \mathbb{P}[\delta \geq k] &
  = \begin{cases} e^{- \frac{e^{-\beta ^*}}{(k-1)!}},
&\textrm{if
}\lim\limits_{n \to \infty} \beta_n = \beta ^* \in (-\infty, \infty), \\
1, &\textrm{if }\lim\limits_{n \to \infty} \beta_n = \infty, \\ 0,
&\textrm{if }\lim\limits_{n \to \infty} \beta_n = -\infty.
\end{cases} \nonumber%\label{eqn_minnodeq10}
 \end{align}

\end{cor}

\begin{rem}
Corollary \ref{cor:exact_qcomposite} presents the asymptotically
exact probability and a zero--one law \cite{yagan} (a kind of phase
transition \cite{phasetransition}) for the event that graph
$\mathbb{G}_q$ has a minimum node degree no less
than $k$.

\end{rem}

\begin{rem}

Setting $p_n$ to $1$ in Theorem \ref{thm:exact_qcomposite2} and
Corollary \ref{cor:exact_qcomposite}, we obtain corresponding
results \cite{qcomp_kcon} for topological properties in uniform
$q$-intersection graph $G_q(n, K_n, P_n)$.

\end{rem}

\begin{rem}

In Theorem \ref{thm:exact_qcomposite2} and Corollary
\ref{cor:exact_qcomposite}, given $ K_n = \omega(1)$, we have $q <
K_n$ for all $n$ sufficiently large since $q$ does not scale with
$n$. From $\frac{{K_n}^2}{P_n} = o(1)$, it is clear that $K_n < P_n$
for all $n$ sufficiently large.

\end{rem}

%\begin{rem} \label{remq1}
%
%In the case of $q=1$, we have proved the results of Theorem
%\ref{thm:exact_qcomposite2} and Corollary \ref{cor:exact_qcomposite}
%without the condition ${{K_n}^2}/{P_n} = o(1)$, yet under a weaker
%condition: $P_n \geq 3K_n $ for all $n$ sufficiently large. The
%details can be found in our technical report \cite{QcompTech14} and
%is again omitted due to the space limitation.
%
%
%\end{rem}

\subsection{Practicality of Conditions}

We check the practicality of the conditions in Theorem
\ref{thm:exact_qcomposite2} and Corollary
\ref{cor:exact_qcomposite}: $ K_n = \omega(1)$,
$\frac{{K_n}^2}{P_n}=o(1)$, and (\ref{peq1})--(\ref{peq2}). First,
the condition $ K_n = \omega(1)$ follows in wireless sensor network
applications \cite{yagan_onoff} since $ K_n $ is often logarithmic
\cite{yagan_onoff} with $n$, the number of sensor nodes
in the network. Second, the condition $\frac{{K_n}^2}{P_n}=o(1)$
also holds in practice since the key pool size $P_n$ is expected to
be several orders of magnitude larger than $K_n$
\cite{virgil,adrian}. Finally, (\ref{peq1})--(\ref{peq2}) present
the range of $p_{e, q}$ that is of interest.

\subsection{Analogs of Theorem \ref{thm:exact_qcomposite2}
and Corollary \ref{cor:exact_qcomposite}}~

Analogous results to those of Theorem \ref{thm:exact_qcomposite2}
and Corollary \ref{cor:exact_qcomposite} can be given with $p_{e,q}
$ at all places substituted by $p_n \cdot \frac{1}{q!} \big(
\frac{{K_n}^2}{P_n} \big)^{q}$, due to ${p_{e,q}} = p_n \cdot
p_{s,q}$ from (\ref{eq_pre}) and the replacement of $p_{s,q}$ by
$\frac{1}{q!} \big( \frac{{K_n}^2}{P_n} \big)^{q}$ given Lemma
\ref{lem_eval_psq} below. However, \emph{extra} conditions have to
be added for some results. The details as well as the proof of Lemma
\ref{lem_eval_psq} are provided in the full version
\cite{QcompTech14}.

%
% in
%$\mathbb{G}$  a quantity expressed by $K_n, P_n$ and $q$; i.e., with
% replaced , and hence with
%$p_{e,q} $ replaced by  due to   (the conditions Lemma
%\ref{lem_eval_psq} requires hold in both Theorem
%\ref{thm:exact_qcomposite2} and Corollary
%\ref{cor:exact_qcomposite}). Thus, with (\ref{peq1}) (resp.,
%(\ref{peq2})) replaced by $p_n \cdot \frac{1}{q!} \big(
%\frac{{K_n}^2}{P_n} \big)^{q} = \frac{\ln  n \pm O(\ln \ln n)}{n}$
%(resp., $p_n \cdot \frac{1}{q!} \big( \frac{{K_n}^2}{P_n} \big)^{q}
%=
% \frac{\ln  n + {(k-1)} \ln \ln n + {\beta_n}}{n}$), and keeping all the
%conditions in Theorem \ref{thm:exact_qcomposite2} (resp., Corollary
%\ref{cor:exact_qcomposite}), we demonstrate that the properties (a)
%and (b) in Theorem \ref{thm:exact_qcomposite2} (resp.,
%(\ref{eqn_minnodeq10}) in Corollary \ref{cor:exact_qcomposite})
%still hold\footnote{The results still hold if we further replace
%$p_{e,q} $ in (\ref{newpeq})?? with $\frac{1}{q!} \big(
%\frac{{K_n}^2}{P_n} \big)^{q}$.}.

\begin{lem} \label{lem_eval_psq}

If $ K_n = \omega(1)$ and $\frac{{K_n}^2}{P_n} = o(1)$, then it
follow that $p_{s,q} \sim \frac{1}{q!} \big( \frac{{K_n}^2}{P_n}
\big)^{q} $.
\end{lem}

%
%\begin{proof}
%See our technical report \cite{QcompTech14}.
%\end{proof}

 \begin{figure}[!t]
  \centering 
  \vspace{5pt}
 \includegraphics[height=0.27\textwidth]{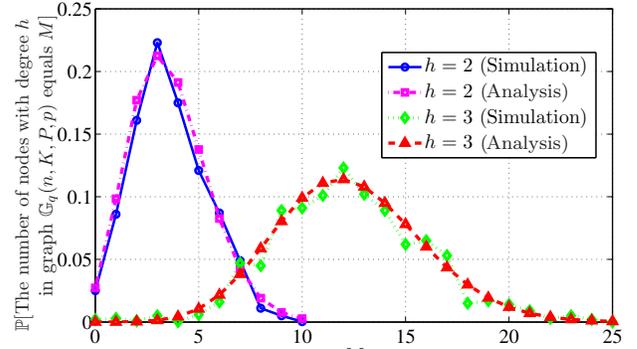}% \vspace{-5pt}
  \vspace{-15pt}\caption{A plot of the probability
 distribution for the number of nodes with
degree $h$ for $h=2,3 $ in graph $\mathbb{G}_q(n,K,P,p)$ with
$n=2,000$, $q = 2$, $P=10,000$, $K = 36 $ and $p = 0.7 $.
\vspace{-5pt}} \label{f4}
\end{figure}

 \begin{figure}[!t]
  \centering
 \includegraphics[height=0.27\textwidth]{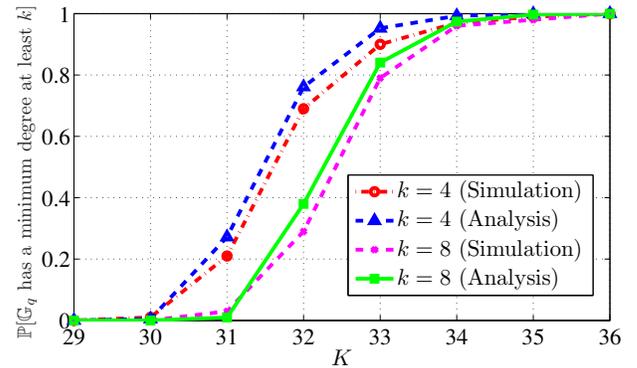}%  \vspace{-5pt}
\caption{A plot of the probability that graph
$\mathbb{G}_q(n,K,P,p)$ has a minimum node degree at least $k$ as a
function of $K$ for
         $k=4$ and $k=8$
         with $q=2$, $n=2,000$, $P=10,000$, and $p=0.8$.
         }
\label{f5}\vspace{-10pt}
\end{figure}

\section{Numerical Experiments} \label{sec:expe}

To confirm the results in Theorem \ref{thm:exact_qcomposite2}, we
now provide numerical experiments in the non-asymptotic regime. As
we will see from the simulation results, the experimental
observations are in agreement with
 our theoretical findings.

In all experiments, we fix the number of nodes at $n=2,000$ and the
key pool size at $P=10,000$. In Figure \ref{f4}, we plot the
probability
 distribution for the number of nodes with degree $h$
  in graph $\mathbb{G}_q(n,K,P,p)$ for $h=2,3$ from both the
simulation and the analysis, with $q = 2$, $K = 36$ and $p = 0.7 $.
On the one hand, for the simulation, we generate $2,000$ independent
samples of $\mathbb{G}_q(n,K,P,p)$ and record the
 count (out of a possible $2,000$) that the number
of nodes with degree $h$ for each $h$ equals a particular
non-negative number $M$ (M is the horizontal axis in Figure
\ref{f4}). Then the empirical probabilities are obtained by dividing
the
 counts by $2,000$. On the other hand, we approximate the analytical curves by the asymptotic
results as explained below. Property (a) of Theorem
\ref{thm:exact_qcomposite2} notes that with the parameter conditions
therein, the number of nodes in
$\mathbb{G}_q(n,K_n,P_n,p_n)$ with degree $h$ is asymptotically equivalent in distribution to a Poisson random variable
with mean $\lambda_{n,h} =n (h!)^{-1}(n p_{e,q})^h e^{-n p_{e,q}}$.
We derive $\lambda_{n,h}$ by computing the corresponding probability of
$p_{e,q}$ in $\mathbb{G}_q(n,K,P,p)$ through 
\begin{align}
p_{e,q} &= p \cdot \sum_{u=q}^{K}
\bigg[{\binom{K}{u}\binom{P-K}{K-u}}\bigg/{\binom{P}{K}}\bigg]
\label{eq:thresholdsab}
\end{align}
 given (\ref{psq1}--\ref{eq_pre}) and $P > 2 K$. Then for each $h$,
we plot a Poisson distribution with mean $\lambda_{n,h}$ as the curve
corresponding to the analysis. We observe that the curves generated
from the simulation and those obtained by the analysis are close to
each other, confirming the result on asymptotic Poisson distribution
in property (a) of Theorem \ref{thm:exact_qcomposite2}.

In Figure \ref{f5}, we depict the probability that graph
$\mathbb{G}_q(n,K,P,p)$ has a minimum node degree at least $k$ from
both the simulation and the analysis, for $k=4$ and $k=8$ with $q =
2$ and $p=0.8$ and $K$ varying from 29 to 36 (we still set $n=2,000$
and $P=10,000$). Similar to the experiments for Figure \ref{f4}
above, we also generate $2,000$ independent samples of graph
$\mathbb{G}_q(n,K,P,p)$ and record the count that the minimum degree
of graph $\mathbb{G}_q(n,K,P,p)$ is no less than $k$; and the
empirical probability of $\mathbb{G}_q(n,K,P,p)$ having a minimum
degree at least $k$ is derived by averaging over the $2,000$
experiments. The analytical curves in Figure \ref{f5} are also
approximated by the asymptotical results as follows. First, we
compute the corresponding probability of $p_{e,q}$ in
$\mathbb{G}_q(n,K,P,p)$ through (\ref{eq:thresholdsab}). Then based
on (\ref{peq2}), we determine $\beta$ through $p_{e, q} = \frac{\ln
n + {(k-1)} \ln \ln n + {\beta}}{n}$. Then with an approximation to
the asymptotical results in Corollary \ref{cor:exact_qcomposite}, we
plot the analytical curves by considering that the minimum degree of
$\mathbb{G}_q(n,K,P,p)$ is at least $k $ with probability $e^{-
\frac{e^{-\beta}}{(k-1)!}}$. The observation that the curves
generated from the simulation and the analytical curves are close to
each other is in accordance with Corollary
\ref{cor:exact_qcomposite}.

\section{Related Work} \label{related}

Erd\H{o}s and R\'{e}nyi \cite{citeulike:4012374} propose the random graph model $G(n,p_n)$ defined on
a node set with size $n$ such that an edge between any two nodes
exists with probability $p_n$ \emph{independently} of all other
edges. For graph $G(n,p_n)$, Erd\H{o}s and R\'{e}nyi
\cite{citeulike:4012374} derive the asymptotically exact
probabilities for connectivity the property that the minimum degree
is at least $1$, by proving first that the number of isolated nodes
converges to a Poisson distribution as $n \to \infty$. Later, they
extend the results to general $k$ in \cite{erdos61conn}, obtaining
the asymptotic Poisson distribution for the number of nodes with certain degree and the asymptotically exact probabilities for
$k$-connectivity and the event that the minimum degree is at least
$k$, where $k$-connectivity is defined as the property that the
network remains connected in spite of the removal of any $(k-1)$
nodes\footnote{$k$-connectivity throughout this paper means
$k$-vertex-connectivity \cite{DBLP:conf/soda/Censor-HillelGK14}.}.
Since its introduction, graph $G(n,p_n)$ has been widely
investigated
\cite{RSA:RSA20342,JZIS14,Pittel2005127,FriezeLoh,Chung2001257,Gamarnik,beer2011vertex-journal,Flaxman,Johansson,Beame,BONATO}.

For graph $G_q(n, K_n, P_n)$, Bloznelis \emph{et al.}
\cite{Rybarczyk} demonstrate that a connected component with at at
least a constant fraction of $n$ emerges asymptotically when the edge
probability $p_{e,q}$ exceeds $1/n$. Bloznelis and {\L}uczak
\cite{Perfectmatchings} have recently considered connectivity and
perfect matching. Still in $G_q(n, K_n, P_n)$, Bloznelis \emph{et
al.} \cite{Assortativity} investigate assortativity and clustering,
while for asymptotic node degree distribution, Bloznelis
\cite{bloznelis2013} analyzes clustering coefficient and the degree
distribution of a typical node. We \cite{qcomp_kcon} compute the
probability distribution for the minimum node degree. Recently, Bloznelis and Rybarczyk \cite{Bloznelis201494} and we \cite{ANALCO} have derived the asymptotically exact probability of
 $k$-connectivity. Several variants or generalizations of
graph ${G}_q(n, K_n, P_n)$ are also considered in the literature
\cite{Rybarczyk,bloznelis2013,Assortativity,DBLP:conf/waw/BradonjicHHP10}.

 When $q=1$, for graph ${G}_1(n, K_n, P_n)$
(also referred to as a random key graph
\cite{ISIT_RKGRGG,yagan,virgillncs} or a uniform random intersection
graph \cite{ryb3,r1}) and some of its variants, a number of
properties have been extensively studied in the literature including
component evolution \cite{DBLP:journals/corr/abs-1301-7320},
connectivity \cite{ryb3,yagan,r1}, $k$-connectivity \cite{zz,JZCDC},
node degree distribution
\cite{Jaworski20062152,BloznelisD13,PES:6114960,RSA:RSA20005} and
independent sets
\cite{Nikoletseas:2008:LIS:1414105.1414429,Rybarczyk2014103}.

In graph $\mathbb{G}_1$,
Ya\u{g}an \cite{yagan_onoff} presents zero--one laws for
connectivity and for the property that the minimum degree is at
least $1$. We extend Ya\u{g}an's results to
general $k$ for $\mathbb{G}_1$ in \cite{ISIT,ZhaoYaganGligor}.

Krishnan \emph{et al.} \cite{ISIT_RKGRGG} and Krzywdzi\'{n}ski and
Rybarczyk \cite{Krzywdzi} describe results for the probability of
connectivity asymptotically converging to 1 in WSNs employing the
$q$-composite key predistribution scheme with $q=1$ (i.e., the basic
Eschenauer--Gligor key predistribution scheme), not under the on/off
channel model but under the well-known disk model
\cite{Saniee,Bradonjic:2010:EBR:1873601.1873715,Gupta98criticalpower,goel2005,Zhao:2011:FRN:2030613.2030651,Mobilegeometric},
where nodes are distributed over a bounded region of a Euclidean
plane, and two nodes have to be within a certain distance for
communication. Simulation results in our work \cite{ZhaoYaganGligor}
indicate that for WSNs under the key predistribution scheme with
$q=1$, when the on-off channel model is replaced by the disk model,
the performances for $k$-connectivity and for the property that the
minimum degree is at least $k$ do not change significantly.

\section{Conclusion and Future Directions}
\label{sec:Conclusion}

In this paper, we analyze several topological properties related to
node degree in a wireless sensor network operating under the
$q$-composite key predistribution scheme with on/off channels. The
network is modeled by the superposition of an Erd\H{o}s-R\'enyi
graph on a uniform $q$-intersection graph. Numerical simulation is
shown to be in agreement with our theoretical findings.

Two future research directions are as follows.  To begin with, we
can consider physical link constraints different with the on/off
channel model, where one candidate is the aforementioned disk model.
Another extension is to derive the asymptotically exact probability
and thus a zero--one law for $k$-connectivity in graph
$\mathbb{G}_q$. Note that a zero--law for $k$-connectivity follows
immediately from Corollary \ref{cor:exact_qcomposite} since
$k$-connectivity implies the property of minimum node degree being
at least $k$. The one--law and the asymptotically exact probability
result will follow if we show that under certain conditions, the
probability that $\mathbb{G}_q$ has a minimum node degree no less
than $k$ but is not $k$-connected converges to $0$ as $n \to
\infty$.

\end{document}